\documentclass{article}
\usepackage{spconf,amsmath,graphicx}
\usepackage{hyperref}
\usepackage{color}
\usepackage{amssymb}
\usepackage{booktabs}
\usepackage{setspace}

\title{CrossSinger: A Cross-Lingual Multi-Singer High-Fidelity Singing Voice Synthesizer Trained on Monolingual Singers}
\name{$^*${Xintong Wang}$^1$, $^*${Chang Zeng}$^{2,3}$, Jun Chen$^4$, Chunhui Wang$^1$}
\address{$^1$Beijing Bombax XiaoIce Technology Co., Ltd, China \\ $^2$National Institute of Informatics, Japan $^3$SOKENDAI, Japan \\ $^4$Shenzhen International Graduate School, Tsinghua University, Shenzhen, China}
\begin{document}
\ninept
\maketitle
\renewcommand{\thefootnote}{\fnsymbol{footnote}}
\footnotetext[1]{These authors contributed equally to this work.}
%
\begin{abstract}
It is challenging to build a multi-singer high-fidelity singing voice synthesis system with cross-lingual ability by only using monolingual singers in the training stage. In this paper, we propose CrossSinger, which is a cross-lingual singing voice synthesizer based on Xiaoicesing2. Specifically, we utilize International Phonetic Alphabet to unify the representation for all languages of the training data. Moreover, we leverage conditional layer normalization to incorporate the language information into the model for better pronunciation when singers meet unseen languages. Additionally, gradient reversal layer (GRL) is utilized to remove singer biases included in lyrics since all singers are monolingual, which indicates singer's identity is implicitly associated with the text. The experiment is conducted on a combination of three singing voice datasets containing Japanese Kiritan dataset, English NUS-48E dataset, and one internal Chinese dataset. The result shows CrossSinger can synthesize high-fidelity songs for various singers with cross-lingual ability, including code-switch cases.
\end{abstract}
\section{Introduction}
\label{sec:intro}
With the advancements in deep learning and the continuous evolution of computing resources, speech synthesis has experienced significant progress, especially in acoustic models \cite{tacotron2, gantts, wgantts, fs1, fs2} and vocoders \cite{wavenet, nsf, hifigan}. These two key components play crucial roles in modern speech synthesis systems. Given the similarity in workflow between speech synthesis and singing voice synthesis (SVS), the recent breakthroughs in the speech synthesis community \cite{48khz1, lee-koreasvs, xiaoice} have also greatly benefited the field of SVS. Consequently, SVS has gained considerable attention from both academia and industry.
Notably, the architectural framework of Fastspeech2 \cite{fs2} has been adopted in singing voice synthesis models such as Xiaoicesing \cite{xiaoice} and its improved version, Xiaoicesing2 \cite{xiaoice2}. These models have demonstrated the capability to generate high-fidelity singing voices. Additionally, SingGAN \cite{singgan} effectively employed the sine excitation method originally proposed in \cite{nsf} to synthesize high-quality singing voices.

The cross-lingual scenario is a pivotal topic within the speech synthesis community, given the prevalence of multilingual speakers in today's world \cite{cltts-liming}. Numerous studies \cite{cltts-chenmengnan, cltts-zhangyu, cltts-zhouxuehao} have focused on addressing this aspect. However, when it comes to SVS, the cross-lingual scenario remains relatively unexplored. In reality, most professional singers are proficient in only one language, which restricts many foreign fans from experiencing their unique timbres and expressive styles in different languages. Consequently, the development of a high-fidelity singing voice system with cross-lingual capabilities is essential to cater to their diverse needs, achieved through training on monolingual singers.


To this end, in this paper, we extend the monolingual high-fidelity SVS system Xiaoicesing2 \cite{xiaoice2} to a cross-lingual multi-singer singing voice synthesizer, CrossSinger, by utilizing a unified representation space and incorporating language information into the model. Specifically, we first unify the phonetic representation for all languages by using the International Phonetic Alphabet (IPA). In this way, all languages can share the same representation space, even if they have different grapheme sets and pronunciations \cite{cltts-liming}. Additionally, we incorporate the language information into the model to enable it to learn the intrinsic rules and grammar for each language during the training stage. To achieve this, we employ conditional layer normalization (CLN) \cite{adaspeech} to fuse the language information when synthesizing the corresponding mel-spectrogram.
Furthermore, since all singers in our training dataset are monolingual singers, their identities become implicitly associated with the input lyrics. This association can introduce biases when the model learns from this data without special consideration. To address this issue, we introduce a singer bias eliminator, comprising a singer classifier and a gradient reversal layer \cite{grl}, to remove singer-specific information from the lyrics.

In the experiment, we train CrossSinger using a combination of three datasets: the Japanese Kiritan \cite{kiritan} dataset, the English NUS-48E \cite{nus-48e} dataset, and a single-singer Chinese dataset. The experimental results demonstrate that our CrossSinger can generate singing voices with high naturalness and intelligibility in the cross-lingual scenario, including code-switch cases.

The remainder of this paper is organized as follows. In Section \ref{sec:xiaoice2}, we provide a brief review of Xiaoicesing2 \cite{xiaoice2}, as our CrossSinger model is built upon it. The detailed architecture and methodology of CrossSinger are presented in Section \ref{sec:crosssinger}. In Section \ref{sec:exps}, we showcase the experimental results\footnote{\href{https://wavelandspeech.github.io/CrossSinger}{\text{Demo page: https://wavelandspeech.github.io/CrossSinger}}}. Finally, we conclude this paper in Section \ref{sec:conclusion}, summarizing the key findings and discussing potential future directions.


\section{Review of Xiaoicesing2}
\label{sec:xiaoice2}

Xiaoicesing2 \cite{xiaoice2} is a state-of-the-art monolingual high-fidelity singing voice synthesizer that leverages generative adversarial networks (GAN) \cite{gan}. The model consists of a generator, adapted from Fastspeech2 \cite{fs2}, and a multi-band discriminator similar to HiFiSinger \cite{hifisinger}. Notably, the authors enhance the architecture of the feed-forward Transformer (FFT) block \cite{s2sfft} in the generator by integrating multiple residual $1$-d convolutional blocks in parallel with the multi-head attention module to fuse global and local features effectively \cite{xiaoice2}. The training process involves combining acoustic loss \cite{fs2} and feature match loss \cite{melgan} with the adversarial loss proposed in LS-GAN \cite{lsgan} to train the generator effectively.

Regarding the discriminator, due to the demanding nature of high-fidelity SVS at a high sampling rate (e.g., $48$kHz), which requires generating high-resolution mel-spectrograms with frequency ranges exceeding $20$kHz, the authors employ a multi-discriminator approach. They divide the entire mel-spectrogram into multiple sub-band areas and apply a dedicated discriminator for each sub-band mel-spectrogram. Specifically, a group of segment discriminators \cite{hifisinger} and detail discriminators \cite{patchgan1, patchgan2, patchgan3} are combined to accurately differentiate between real audio and generated sub-band mel-spectrograms.

Experimental evaluation conducted on an internal single-singer Chinese dataset showcases the remarkable capability of Xiaoicesing2 in generating high-quality singing voices that rival human performance. Additionally, the ablation study presented in \cite{xiaoice2} highlights the significant improvement in MOS score achieved through adversarial training when compared to the counterpart without discriminators.

\section{CrossSinger}
\label{sec:crosssinger}

\begin{figure}
    \centering
    \includegraphics[width=8cm]{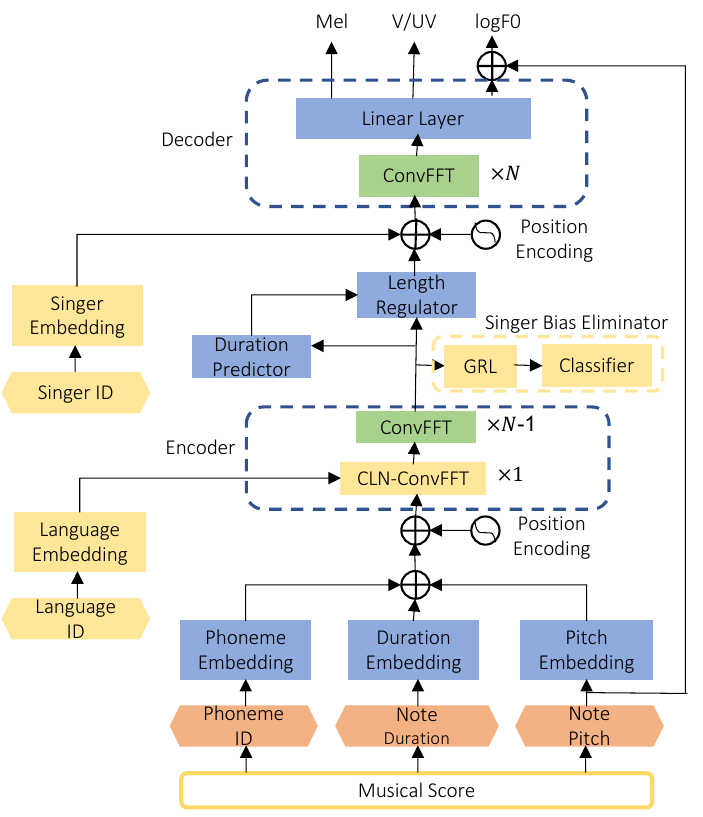}
    \caption{The architecture of the CrossSinger generator. Yellow blocks denote the improvement parts compared with Xiaoicesing2.}
    \label{fig:generator}
\end{figure}

While Xiaoicesing2 has demonstrated its ability to synthesize high-fidelity singing voices, it falls short in handling the cross-lingual scenario, where the model needs to generate singing voices based on foreign lyrics for a monolingual singer \cite{cltts-liming}. To overcome this limitation, we propose an innovative approach that incorporates language information and singers' identities into Xiaoicesing2, giving rise to a novel cross-lingual multi-singer singing voice synthesizer named CrossSinger. The architecture of CrossSinger is depicted in Figure \ref{fig:generator}, with our contributions represented by the highlighted yellow blocks. For conciseness, we showcase only the updated generator of CrossSinger in this figure, as the discriminator remains unchanged. Subsequently, in the forthcoming sections, we delve into the comprehensive explanation of CrossSinger, illuminating how it effectively learns cross-lingual capabilities through training solely on monolingual singers.

\subsection{Phonetic representation}
\label{sec:ipa}
One of the primary challenges in constructing a cross-lingual speech or singing voice synthesizer is achieving a unified representation for all languages, a topic that has been explored in speech synthesis studies \cite{ipa1, ipa2, bytes}. In this work, we adopt the approach presented in \cite{ipa2}, utilizing the International Phonetic Alphabet (IPA) to represent all phonemes across languages. As illustrated in Figure \ref{fig:generator}, the musical score is parsed into three sequences: the phoneme sequence, note duration sequence, and note pitch sequence. Similar to the transformations in Xiaoicesing2, where these sequences are mapped to individual embedding spaces, we perform similar operations in CrossSinger. However, the phoneme set used in CrossSinger is larger compared to Xiaoicesing2, given the inclusion of multiple languages. Furthermore, the employment of a unified phonetic representation space allows for similar pronunciations across different languages to be denoted by a single phoneme, which proves to be advantageous when enabling monolingual singers to sing in foreign languages.

\subsection{Language conditional layer normalization}
\label{sec:cln}

\begin{figure}
    \centering
    \includegraphics[width=8cm]{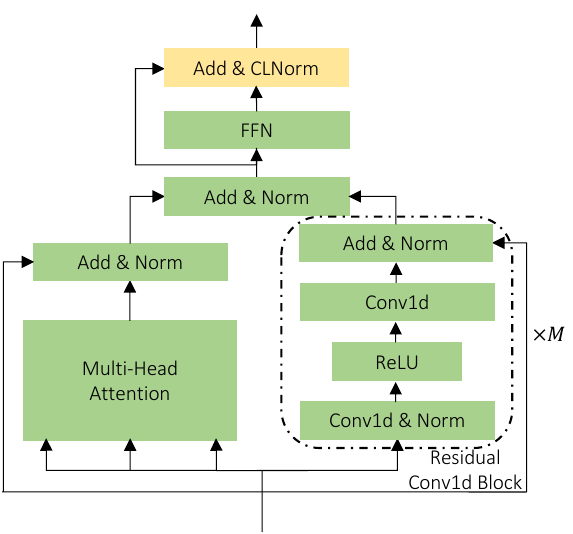}
    \caption{The architecture of ConvFFT block with conditional layer normalization. CLN is used to replace the last layer normalization of ConvFFT to introduce the language information.}
    \label{fig:cln_convfft}
\end{figure}

In addition to adopting a unified phonetic representation, explicitly incorporating language information proves to be beneficial in learning the individual grammar and rules of each language. To achieve this, we first transform the language information into a language embedding space. Subsequently, the language embedding is fused with the output of the feed-forward network (FFN) within the first ConvFFT block, as depicted in Figure \ref{fig:cln_convfft}. This fusion is achieved through Conditional Layer Normalization (CLN) \cite{adaspeech}, involving two affine layers that transform the language embedding into the scale ($\boldsymbol{\alpha}$) and bias ($\boldsymbol{\beta}$) parameters, respectively. These scale and bias parameters are then applied to the output of the FFN. Mathematically, this can be formulated as follows:
\begin{align}
\label{eq:cln}
\boldsymbol{\alpha} & = \boldsymbol{W}_{\alpha}^\top \cdot \boldsymbol{e}^l, \\
\boldsymbol{\beta} & = \boldsymbol{W}_{\beta}^\top \cdot \boldsymbol{e}^l, \\
\text{CLN}(\boldsymbol{X}) & = \boldsymbol{\alpha}\odot\frac{\boldsymbol{X} - \boldsymbol{\mu}}{\boldsymbol{\sigma}} + \boldsymbol{\beta},
\end{align}
where $\boldsymbol{e}^l$ represents the language embedding\footnote{In this paper, all vectors are assumed to be column vectors.}, while $\boldsymbol{W}_{\alpha}$ and $\boldsymbol{W}_{\beta}$ denote the parameters of the affine layers for scale $\boldsymbol{\alpha}$ and bias $\boldsymbol{\beta}$, respectively. Additionally, $\boldsymbol{\mu}$ and $\boldsymbol{\sigma}$ represent the mean and standard deviation of the input $\boldsymbol{X}$. Notably, CLN acts on each element of $\boldsymbol{X}$, making it more efficient and flexible compared to simply concatenating the language embedding with $\boldsymbol{X}$\footnote{$\odot$ denotes element-wise multiplication.}. It is important to mention that we only utilize CLN in the first ConvFFT block of the encoder since adding more CLN blocks does not contribute significantly to the quality of the synthesized singing voices in our experiments.

\subsection{Singer bias eliminator}
\label{sec:grl}

\begin{figure}
    \centering
    \includegraphics[width=3cm]{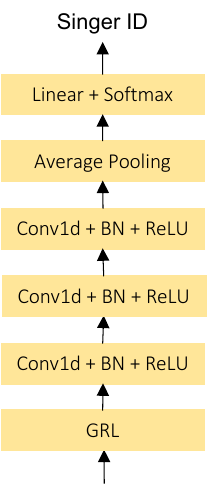}
    \caption{The architecture of singer bias eliminator. It is utilized to remove singer biases implicitly associated with lyrics.}
    \label{fig:eliminator}
\end{figure}

In our training data, all singers are monolingual, and as a result, the identity of the singer becomes implicitly associated with the language of the input lyrics. This association may inadvertently cause the model to learn unexpected biases between singers and languages if left unaddressed. For instance, there could be phonemes that are common in Chinese but rare in Japanese, leading Japanese singers to unnaturally pronounce these Chinese phonemes in a Japanese style due to the unintended leakage of the singer's identity caused by this association.

To overcome such biases, we implement a singer bias eliminator applied to the output of the encoder, as depicted in Figure \ref{fig:generator}. The detailed architecture of the singer bias eliminator is illustrated in Figure \ref{fig:eliminator}. It comprises a gradient reversal layer (GRL) \cite{grl} and a singer classifier. The singer classifier consists of three $1$-d convolutional layers for feature extraction, followed by an average pooling layer for aggregation, and a linear layer with a softmax function for singer classification.

The transformation process of the singer bias eliminator can be formulated as follows:
\begin{align}
\boldsymbol{e}^{s} & = f_{\boldsymbol{\theta}}(\boldsymbol{X}^e) \\
P(\boldsymbol{e}^{s}, i) & = \frac{\text{exp}(\boldsymbol{w}_i^{\top}\cdot\boldsymbol{e}^{s})}{\sum_{j=1}^{S}\text{exp}(\boldsymbol{w}_j^{\top}\cdot\boldsymbol{e}^{s})},
\label{eq:singer}
\end{align}
where $\boldsymbol{X}^{e}$ represents the output of the encoder. $f_{\boldsymbol{\theta}}$ denotes the transformation with learnable parameters $\boldsymbol{\theta}$ of GRL, convolutional layers, and the average pooling layer, whose output is denoted as the speaker embedding $\boldsymbol{e}^s$. $P(\boldsymbol{e}^s, i)$ represents the probability that the speaker embedding $\boldsymbol{e}^s$ belongs to the $i$-th speaker, corresponding to the linear layer with learnable parameters $\boldsymbol{W}=[\boldsymbol{w}_1,\boldsymbol{w}_2,...,\boldsymbol{w}_S]$ and softmax function shown in Fig. \ref{fig:eliminator}. Here, $S$ denotes the number of singers in the training dataset. Due to GRL, the gradient from the classifier is reversed, effectively removing the singer's information implicitly associated with the lyrics. This process proves beneficial in improving the naturalness when pronouncing foreign languages for the singer.

\setlength{\tabcolsep}{4mm}
\begin{table*}[h]
  \caption{MOS test result with $95$\% confidence interval for $48$kHz singing voice synthesis. We evaluate the performance in three aspects, including sound quality, pronunciation accuracy, and naturalness.}
  \label{tab:main_result}
  \centering
  \vspace{4mm}
  \begin{tabular}{lccc}
    \toprule
    \textbf{Systems} & \textbf{Sound Quality}($\uparrow$) & \textbf{Pronunciation Accuracy}($\uparrow$) & \textbf{Naturalness}($\uparrow$) \\
    \midrule
    Ground truth                           & $4.54\pm 0.034$              & $4.78\pm 0.033$          & $4.77\pm 0.033$   \\
    XiaoiceSing2 + HiFi-WaveGAN            & $3.28\pm 0.075$              & $3.32\pm 0.072$          & $3.16\pm 0.077$   \\
    CrossSinger + HiFi-WaveGAN             & $\textbf{4.15}\pm 0.052$     & $\textbf{4.42}\pm 0.047$ & $\textbf{3.98}\pm 0.058$   \\
    \bottomrule
  \end{tabular}
\end{table*}

\setlength{\tabcolsep}{4mm}
\begin{table*}[t]
  \caption{MOS test result with $95$\% confidence interval for the ablation study.}
  \label{tab:ablation}
  \centering
  \vspace{2mm}
  \begin{tabular}{lccc}
    \toprule
    \textbf{Systems} & \textbf{Sound Quality}($\uparrow$) & \textbf{Pronunciation Accuracy}($\uparrow$) & \textbf{Naturalness}($\uparrow$) \\
    \midrule
    Ground truth                           & $4.54\pm 0.034$     & $4.78\pm 0.033$ & $4.77\pm 0.033$   \\
    CrossSinger + HiFi-WaveGAN             & $4.15\pm 0.052$     & $4.42\pm 0.047$ & $3.98\pm 0.058$   \\
    - CLN                                  & $3.94\pm 0.058$     & $3.77\pm 0.069$ & $3.62\pm 0.061$   \\
    - singer bias eliminator               & $4.02\pm 0.055$     & $4.10\pm 0.044$ & $3.42\pm 0.065$   \\
    \bottomrule
  \end{tabular}
\end{table*}

\subsection{Loss function}
In addition to the adversarial loss $\mathcal{L}_{adv}$, acoustic loss $\mathcal{L}_a$, and feature match loss $\mathcal{L}_f$ described in \cite{xiaoice2}, we incorporate an additional singer classification loss, represented by Eq. \ref{eq:classification}, to train the generator.
\begin{align}
\label{eq:classification}
\mathcal{L}_{s} & = -\frac{1}{N}\sum_{n}^{N}\sum_{i=1}^{S}\mathcal{I}(y_{ni} = 1)\log P(\boldsymbol{e}_n^s, i),
\end{align}
where $N$ represents the number of samples in the training dataset. The function $\mathcal{I}(\cdot)$ acts as an indicator function, returning $1$ when the condition is true, and $0$ otherwise. Moreover, $y_{ni}$ denotes the ground truth of the singer's identity.

\section{Experiments}
\label{sec:exps}

\subsection{Dataset}
To comprehensively evaluate the effectiveness of our proposed model, we conducted a series of experiments using a combination of three datasets: a Japanese dataset called Kiritan \cite{kiritan}, an English dataset NUS-48E \cite{nus-48e}, and an internal Chinese dataset. The Japanese and Chinese datasets are both single-singer datasets, with the former comprising $2.11$ hours of singing voices and the latter containing $8.95$ hours of singing data. On the other hand, the English dataset is a multi-singer dataset with a total of $12$ singers and approximately $4.74$ hours of singing voices.

For our experiments, we divided all datasets into training, development, and test subsets, with an allocation ratio of $80\%$, $10\%$, and $10\%$, respectively, for training and evaluating CrossSinger. While the training dataset includes all singers from all three languages, the development and test datasets were limited to only five singers, encompassing the Japanese singer, the Chinese singer, and three English singers.

In the evaluation stage, we synthesized singing voices in all three languages for each of the selected singers, allowing us to conduct comprehensive listening tests and assess the cross-lingual performance of CrossSinger.

\subsection{Experimental settings}
To demonstrate the cross-lingual ability of our proposed model, CrossSinger, we employ Xiaoicesing2 as the baseline system. Both Xiaoicesing2 and CrossSinger are trained using the same optimization strategy outlined in \cite{xiaoice2}. Specifically, we set the batch size to $32$ and train these models on $4$ NVIDIA V100 GPUs for $300$ epochs. For the optimization process, we use Adam \cite{adam} with an initial learning rate of $0.01$, $\beta_{1}$ of $0.9$, $\beta_2$ of $0.98$, and $\epsilon$ of $10^{-9}$ for both generators and discriminators.

Given the challenges associated with training the Transformer \cite{attention-all-you-need} and GAN model, we implement a warmup strategy to adapt the learning rate during the training stage. Initially, the learning rate is set to $0.0001$ and gradually increased to $0.01$ over the first $5000$ training steps. Subsequently, the learning rate is decayed by $0.99$ after each epoch to stabilize the training process.

Furthermore, to achieve high-quality results, both Xiaoicesing2 and CrossSinger are combined with a vocoder called HiFi-WaveGAN \cite{hwg}, which is trained using the strategy described in \cite{hwg}.


\subsection{Subjective evaluation}
We conducted a comprehensive series of experiments to evaluate the performance of our proposed model, using the average Mean Opinion Score (MOS) as the evaluation metric. The evaluation was conducted from three key aspects: sound quality, pronunciation accuracy, and naturalness. Each sub-experiment involved $20$ segments for each language per singer. We recruited a total of $60$ listeners, with $20$ assigned to each language.

The experimental results are presented in Table \ref{tab:main_result}, clearly indicating that CrossSinger significantly outperforms Xiaoicesing2 in terms of cross-lingual ability. Specifically, in the aspect of sound quality, CrossSinger surpassed Xiaoicesing2 by a margin of $0.87$, demonstrating its superior robustness in cross-lingual Singing Voice Synthesis (SVS) scenarios. Moreover, a noticeable difference can be observed in the pronunciation accuracy MOS scores, where CrossSinger achieved a score of $4.42$, while Xiaoicesing2 received a lower score of $3.32$. This improvement can be attributed to the successful incorporation of language information into the model, enabling it to produce more intelligible and accurate pronunciations.

Lastly, our proposed singer bias eliminator contributed significantly to the enhanced naturalness of the synthesized singing voices. This is evident from CrossSinger's higher MOS score for naturalness compared to Xiaoicesing2.

Overall, the experimental results highlight the effectiveness of CrossSinger in achieving high-quality cross-lingual singing voice synthesis, making it a promising advancement over Xiaoicesing2.

\subsection{Ablation study}
In addition to the main experiment explained in the previous section, we also conducted an ablation study to determine the contribution of the proposed components to the overall improvement. For the ablation study, we individually removed the language conditional layer normalization (CLN) and the singer bias eliminator, as indicated in Table \ref{tab:ablation}.

From the table, we observe that both the CLN and singer bias eliminator significantly contribute to the effectiveness of CrossSinger. When the CLN component is removed, the MOS scores for sound quality and naturalness exhibit a slight decrease compared to CrossSinger. However, the MOS score for pronunciation accuracy experiences a significant decline. This outcome suggests that while the CLN component has a relatively minor impact on sound quality and naturalness, it plays a crucial role in enhancing intelligibility, especially in the cross-lingual scenario.

On the other hand, removing the singer bias eliminator demonstrates a similar trend in the sound quality term as observed with the CLN component. Conversely, it has a modest effect on intelligibility while considerably impacting naturalness, resulting in a reduction of the MOS score from $3.98$ to $3.42$.

\section{Conclusion}
\label{sec:conclusion}
In this paper, we present an enhanced version of the Xiaoicesing2 model, called CrossSinger, which is designed for cross-lingual multi-singer singing voice synthesis. To effectively incorporate language information into the original Xiaoicesing2 model, we first standardize the speech representation using IPA annotation. Subsequently, we explicitly integrate the language information by fusing it with the feature map in the Fastspeech2 encoder through conditional layer normalization. Additionally, to overcome the limitations of available data, we introduce a singer bias eliminator that implicitly removes the singer-specific information associated with the lyrics, resulting in a more natural and expressive generated singing voice. The experimental results demonstrate that CrossSinger outperforms Xiaoicesing2 by synthesizing high-quality singing voices with improved pronunciation accuracy and enhanced naturalness. Moreover, the ablation study illustrates the individual contributions of each proposed component.


\bibliographystyle{IEEEbib}
\bibliography{strings,refs}

\end{document}